\documentclass[a4paper,11pt]{article}

\usepackage[english]{babel}
\usepackage{setspace}

\usepackage[a4paper,tmargin=3.5truecm,bmargin=4truecm,rmargin=2.5truecm,
lmargin=3.2truecm,twoside,verbose=true]{geometry}
\usepackage{cancel,graphicx}
\usepackage{hyperref}
\usepackage{amsmath,amssymb,slashed}

\usepackage{amsmath,amssymb}
\usepackage[amsmath, hyperref, thmmarks]{ntheorem}
\numberwithin{equation}{section}

\allowdisplaybreaks[1]
\newcommand\bbone{{\mathbb{I}}}

\renewenvironment{thebibliography}[1]
         {\section*{References}\frenchspacing\small
          \begin{list}{[\arabic{enumi}]}
         {\usecounter{enumi}\parsep=2pt\topsep 0pt
         \settowidth{\labelwidth}{[#1]}
         \leftmargin=\labelwidth\advance\leftmargin\labelsep
         \rightmargin=0pt\itemsep=1pt\sloppy}}{\end{list}}

\theoremsymbol{}
\theorembodyfont{\slshape}
\theoremheaderfont{\normalfont\bfseries}
\theoremseparator{}

\theorembodyfont{\upshape}
\theoremsymbol{\ensuremath{\blacklozenge}}

\theoremstyle{nonumberplain}
\theoremheaderfont{\scshape}
\theorembodyfont{\normalfont}
\theoremsymbol{\ensuremath{\blacksquare}}

\qedsymbol{\ensuremath{_\blacksquare}}
\theoremclass{LaTeX}

\title{Quantum instability of gauge theories on $\kappa$-Minkowski space}
\author{ Kilian Hersent$^a$, Philippe Mathieu$^b$, Jean-Christophe Wallet$^a$}

\begin{document}

\date{}
\maketitle
\vspace*{-1cm}

\begin{center}

\textit{$^a$IJCLab, Universit\'e Paris-Saclay, CNRS/IN2P3, 91405 Orsay, France.}  \\
\textit{$^b$Department of Mathematics, 
University of Notre Dame, Notre Dame, IN 46556, USA.} \\
\textit{}\\
\bigskip
 e-mail:
\texttt{ kilian.hersent@universite-paris-saclay.fr, pmathieu@nd.edu, jean-christophe.wallet@universite-paris-saclay.fr}\\[1ex]

\end{center}
\begin{abstract}
We consider a gauge theory on the 5-d $\kappa$-Minkowski which can be viewed as the noncommutative analog of a $U(1)$ gauge theory. We show that the Hermiticity condition obeyed by the gauge potential $A_\mu$ is necessarily twisted. Performing a BRST gauge-fixing with a Lorentz-type gauge, we carry out  a first exploration of the one loop quantum properties of this gauge theory. We find that the gauge-fixed theory gives rise to a non-vanishing tadpole for the time component of the gauge potential, while there is no non-vanishing tadpole 1-point function for the spatial components of $A_\mu$. This signals that the classical vacuum of the theory is not stable against quantum fluctuations. Possible consequences regarding the symmetries of the gauge model and the fate of the tadpole in other gauges of non-covariant type are discussed.
\end{abstract}


\vfill\eject
\section{Introduction.}\label{section1}

It is currently believed that noncommutative structures are likely to show up at the Planck scale \cite{dopplic, dopplic1}, where Quantum Gravity effects become sizeable \cite{amelin}-\cite{cenous1} hence rendering questionable the standard description of space-time as a smooth manifold to reconcile quantum mechanics and gravity. Among the noncommutative (quantum) spaces considered so far, the $\kappa$-Minkowski space \cite{luk1}-\cite{majid-ruegg} has received a considerable attention for more than two decades \cite{luk2} as it appears to be a good candidate for a quantum space underlying the description of Quantum Gravity at least in some regime \cite{amelin}-\cite{cenous1}. Numerous related works focus on algebraic aspects \cite{luk2} while more phenomenological investigations resulted in a huge literature in connection in particular with Doubly Special Relativity \cite{ame-ca1}-\cite{AC2}, Relative Locality \cite{reloc}-\cite{reloc2} or with the quest of possible detectable effects \cite{amelin}-\cite{cenous1}. The simplest description of the ($d$-dimensional) $\kappa$-Minkowski space $\mathcal{M}_\kappa^d$ is to view it as the enveloping algebra of the Lie algebra of noncommutative coordinates defined by $[x_0,x_i]=\frac{i}{\kappa}x_i$, $[x_i,x_j]=0$, for $i,j=1,\cdots, (d-1)$, where $\kappa$ is the deformation parameter with mass dimension $1$, often identified with the Planck mass in 4 dimensions. The bicrossproduct structure of the $\kappa$-Poincar\'e Hopf algebra $\mathcal{P}^d_\kappa$ exhibits clearly its role as coding the quantum symmetries of $\mathcal{M}_\kappa^d$, as first pointed out in \cite{majid-ruegg} , this latter being the dual of the subalgebra of $\mathcal{P}^d_\kappa$ generated by the so-called deformed translations.\\

$\mathcal{M}_\kappa^d$ can be conveniently modeled \cite{DS}, \cite{PW2018}, by an associative $^*$-algebra equipped with the following star-product and involution{\footnote{Conventions: Latin indices $i,j,\hdots=1,2,...,(d-1)$ refer to space coordinates. Timelike quantities are indexed by $0$. We set $x:=(x_\mu)=(x_0,\vec{x})$, $x.y:=x_\mu y^\mu=x_0y_0+\vec{x}\vec{y}$. The Fourier transform of $f\in L^1(\mathbb{R}^d)$ is defined by $(\mathcal{F}f)(p):=\int d^dx\ e^{-i(p_0x_0+\vec{p}.\vec{x})}f(x)$ with inverse $\mathcal{F}^{-1}$.}}
\begin{align}
(f\star g)(x)&=\int \frac{dp^0}{2\pi}dy_0\ e^{-iy_0p^0}f(x_0+y_0,\vec{x})g(x_0,e^{-p^0/\kappa}\vec{x})  \label{starpro-4d},\\
f^\dag(x)&= \int \frac{dp^0}{2\pi}dy_0\ e^{-iy_0p^0}{\bar{f}}(x_0+y_0,e^{-p^0/\kappa}\vec{x})\label{invol-4d},
\end{align}
for any functions $f,g$ in a suitable multiplier algebra of the algebra of Schwartz functions in $\mathbb{R}^d$ ($\bar{f}$ is the complex conjugate of $f$.). It will be also denoted hereafter by $\mathcal{M}_\kappa^d$.{\footnote{This algebra involves the smooth functions with polynomial bounds as well as all their derivatives, such that their inverse Fourier transform has compact support in the $x_0$ direction \cite{DS}.}}. The construction basically combines the essential features of the Weyl-Wigner quantization map with properties of the convolution algebra of the affine group $\mathbb{R}\ltimes\mathbb{R}^{(d-1)}$. It is nothing but the extension of the old construction related to the Moyal product which used the notion of twisted convolution related to the Heisenberg group \cite{vonNeum} (instead of the affine group) aiming to rigorously formalize the correspondence between classical and quantum observables initially introduced by Weyl \cite{Weyl} throught the (Weyl-Wigner) quantization map. Notice that a somewhat similar extension to the case of $\mathbb{R}^3_\lambda$, a deformation of $\mathbb{R}^3$, based however on the SU(2) group algebra, underlies the studies of quantum properties of field theories on this latter quantum space \cite{Vitale:2012dz}, \cite{Gere:2015ota}, \cite{Wallet:2016ilh}.\\

The star-product \eqref{starpro-4d} is especially convenient for practical purpose, a fact exploited in recent explorations of quantum properties of $\kappa$-Poincar\'e invariant scalar field theories on $\mathcal{M}_\kappa^4$ \cite{PW2018}, \cite{Juric:2018qdi}, \cite{PW2018bis}. The $\kappa$-deformed relativistic symmetries of $\mathcal{M}_\kappa^d$ can be actually viewed as coded in the $\kappa$-Poincar\'e algebra. Hence, requiring $\kappa$-Poincar\'e invariance is physically natural. It is known \cite{DS}, \cite{PW2018}, that this invariance is achieved provided the action is of the form $\int d^dx\ \mathcal{L}$ where $\mathcal{L}$ is some Lagrangian density and $\int d^dx$ is the simple Lebesgue measure which however is no longer cyclic w.r.t. the star-product \eqref{starpro-4d}, since one has for any $f,g\in\mathcal{M}_\kappa^d$
\begin{equation}
\int d^dx\ (f\star g)(x)=\int d^dx\ \left((\mathcal{E}^{d-1}\triangleright g)\star f\right)(x),\label{twistrace}
\end{equation}
where $\mathcal{E}^{d-1}$, called the modular twist, is defined by
\begin{equation}
(\mathcal{E}\triangleright f)(x)=f(x_0+\frac{i}{\kappa},\vec{x}), \ \ \mathcal{E}=e^{-P_0/\kappa},\label{modular-twist-d}
\end{equation}
with $(P_\mu\triangleright f)(x)=-i\partial_\mu f(x)$, $\mu=0,...,d-1$. Eqn. \eqref{twistrace} defines a twisted trace with respect to \eqref{starpro-4d} which, as pointed out in \cite{PW2018}, \cite{matas}, \cite{matas1}, promotes the action to the status of KMS weight \cite{kuster}, \cite{kuster1} a new property which replaces the lost cyclicity{\footnote{KMS weights which are basically KMS states up to a normalization, are linked to the Tomita group of modular automorphisms \cite{takesaki} whose generator here is $\mathcal{E}^{d-1}$. For discussions on physical consequences of KMS property, see \cite{ConRove}.}}. The loss of cyclicity complicates {\it{a priori}} the construction of a suitable gauge-invariant action for a $\kappa$-Poincar\'e invariant gauge theory, starting from any standard (untwisted) noncommutative differential calculus. Indeed, the usual compensation between unitary gauge factors in the noncommutative gauge transformation of the curvature $F_{\mu\nu}$, which takes the generic form $F^g_{\mu\nu}=g^\dag\star F_{\mu\nu}\star g $, $g^\dag\star g=g\star g^\dag=1$, cannot occur  \cite{MW20}, due to the twisted trace relation  \eqref{twistrace}. This prevents the construction of a gauge invariant polynomial action depending on the curvature. For a review on earlier algebraic and field theoretic developments on gauge theories on $\kappa$-Minkowski space and related problems, see \cite{balkan}.\\

As shown in \cite{MW20}, the troublesome effect of the modular twist can be neutralized, leading to a $\kappa$-Poincar\'e invariant {\it{and}} gauge invariant action with physically suitable commutative limit. This happens thanks to the existence of a unique twisted noncommutative differential calculus based on a family of twisted derivations related to the deformed translations of the $\kappa$-Poincar\'e algebra  \cite{MW20}. This forces the gauge transformations of the curvature to be twisted thus allowing now the unitary gauge factors to balance each other, which actually occurs for a unique value of $d$, namely $d=5$ \cite{MW20}. This comes from the fact that the modular twist \eqref{modular-twist-d} depends on the dimension $d$. Otherwise stated, the main physical prediction is that the coexistence of $\kappa$-Poincar\'e invariance and gauge invariance implies the existence of one extra dimension. For a first exploration of phenomenological features of the 4-dimensional theory obtained from compactification scenarii, see \cite{Mathieu:2020ywc}.\\

 Notice that the analysis carried out in \cite{MW20} used the notion of noncommutative connection on a right module $\mathbb{E}$. It was assumed to be one copy of the algebra $\mathcal{M}_\kappa^5$ in order to describe a noncommutative analog of a $U(1)$ Yang-Mills theory while the action of the algebra $\mathcal{M}_\kappa^5$ on $\mathbb{E}$ was chosen to be $m\triangleleft a=m\star a$, for any $m\in\mathbb{E}$, $a\in\mathcal{M}_\kappa^5$, owing to $\mathbb{E}\simeq\mathcal{M}_\kappa^5$. One may wonder if a suitable choice for this action would lead to another value for $d$, since the corresponding gauge transformations should be modified. Algebraic constraints from right module structure together with physical requirement lead to actions of the form $m\triangleleft a=m\star \phi(a)$ where $\phi$ is any regular automorphism of the algebra, i.e. $(\phi(a))^\dag=\phi^{-1}(a^\dag)$. Again, one finds \cite{HMW} that $d=5$ is the only allowed value for which $\kappa$-Poincar\'e invariant gauge theories exist. Thus, occurrence of one extra dimension seems to be a rather robust feature of these gauge theories on $\kappa$-Minkowski, which motivates further investigation on their quantum properties. \\

This is the purpose of the present paper which will exploit the BRST symmetry linked to the twisted gauge symmetry elaborated in \cite{Mathieu:2021mxl}. As a first step in the investigation, we will compute the one-loop 1-point (tadpole) function for this 5-dimensional noncommutative gauge theory without matter. We will also pay attention to the Hermiticity condition obeyed by the gauge potential. We find that this latter Hermiticity condition is twisted. Furthermore, we find that the gauge-fixed theory gives rise to a non-vanishing tadpole for the time component of the gauge potential, while there is no non-vanishing tadpole 1-point function for the spatial components of $A_\mu$. This signals that the classical vacuum of the theory is not stable against quantum fluctuations. Some possible consequences from the viewpoint of symmetries, namely Lorentz symmetry and noncommutative gauge symmetry are then discussed.\\

The paper is organized as follows. In subsection \ref{section21}, we first collect the useful properties which will be needed in the course of the analysis, including essential features of the twisted differential calculi singled out in \cite{MW20}, \cite{Mathieu:2020ywc}. We then show that the Hermiticity condition affecting the (noncommutative analog of the) gauge potential $A_\mu$ becomes twisted. In the subsection \ref{section22}, we present the BRST gauge-fixing of the classical action. Section \ref{section3} is centered on the one-loop computation of the tadpole 1-point function for the gauge potential. We find that the ghost-gauge vertex and the trilinear vertex of the gauge-fixed action give non-vanishing  contributions only to the 1-point function for $A_0$, the time component of $A_\mu$. The tadpole function for the spatial components of $A_\mu$ are all zero. In section \ref{section4}, we discuss the results and conclude.

\section{Gauge theories on \texorpdfstring{$\kappa$}{k}-Minkowski space}\label{section2}
\subsection{Basic features of the classical action.}\label{section21}
The classical gauge-invariant action is given by \cite{MW20}
\begin{equation}
S_{cl}=\frac{1}{4}\int d^5x\ (F_{\mu\nu}\star F_{\mu\nu}^\dag)(x)\label{actionclass},
\end{equation}
where the curvature can be expressed as
\begin{equation}
F_{\mu\nu}=\mathcal{E}^{-2\gamma}\triangleright(X_\mu A_\nu-X_\nu A_\mu)+(\mathcal{E}^{1-\gamma}\triangleright A_\mu)\star(\mathcal{E}^{\gamma}\triangleright A_\nu)-(\mathcal{E}^{1-\gamma}\triangleright A_\nu)\star(\mathcal{E}^{\gamma}\triangleright A_\mu)\label{fmunu},
\end{equation}
in which $\gamma$ is a real parameter and $A_\mu$ denotes as usual (the noncommutative analog of) the gauge potential which is assumed in the following to be {\it{real-valued}}. Notice that in 5 dimensions the action \eqref{actionclass} should normally be rescaled by a dimensionful parameter $\frac{1}{g^2}$ where $g^2$ has mass dimension $-1$. As this parameter is not essential in the following discussion, we will omit it. \\
In \eqref{actionclass}, the $X_\mu$'s are the generators of an Abelian Lie algebra of twisted derivations denoted by $\mathfrak{D}_\gamma$. These are built from the so-called deformed translations which generate a sub-Hopf algebra of the $\kappa$-Poincar\'e algebra. The twisted derivations are given by
\begin{equation}
X_0=\kappa\mathcal{E}^\gamma(1-\mathcal{E}),\ \ X_i=\mathcal{E}^\gamma P_i,\ \ i=1,2,3,4,\label{xmiou}
\end{equation}
and satisfy the following twisted Leibniz rule
\begin{equation}
X_\mu(a\star b)=X_\mu(a)\star(\mathcal{E}^\gamma\triangleright b)+(\mathcal{E}^{\gamma+1}\triangleright a)\star X_\mu(b)\label{leibnitz},
\end{equation}
for any elements $a,b$ of $\mathcal{M}_\kappa^5$. The twisted derivations of $\mathfrak{D}_\gamma$ generate the twisted noncommutative differential calculus underlying the analysis. The corresponding relevant properties are collected in the appendix \ref{appendixA}. For general algebraic properties of (untwisted) derivation-based differential calculus, see e.g. \cite{mdv} and references therein. \\
Note that interesting twisted differential calculi for $\kappa$-Minkowski spaces stemming from the use of twist deformation formalism applied to abelian and Jordanian twists are considered and discussed in \cite{balkan}. These are not relevant here. Besides, interesting bicovariant (untwisted) differential 
calculi on $\kappa$-Minkowski spaces \cite{sit, sit22} cannot give rise to $\kappa$-Poincar\'e invariant and gauge invariant action functionals as discussed in \cite{MW20}.\\

In this paper, we will use a twisted version of the notion of noncommutative connection on a right-module over $\mathcal{M}_\kappa^5$, introduced in \cite{MW20}. The main mathematical properties of the notion of untwisted connection on a right(or left-)module are characterized in details in \cite{mdv} and  further developped in \cite{cawa}-\cite{epsilonconnect}). Notice that untwisted connections on right-module underly the pioneering works \cite{JMadore}-\cite{JMadore2}.\\

Recall that the gauge group $\mathcal{U}$ is defined as the set of automorphisms of the right-module over $\mathcal{M}_\kappa^5$, denoted by $\mathbb{E}$, assumed here to be one copy of $\mathcal{M}_\kappa^5$, i.e. $\mathbb{E}\simeq\mathcal{M}_\kappa^5$, which are required to preserve the canonical hermitian structure
\begin{equation}
h(m_1,m_2)=m_1^\dag\star m_2
\end{equation}
for any $m_1,m_2\in\mathcal{M}_\kappa^5$. It is a simple matter of algebra to find that
\begin{equation}
\mathcal{U}=\{ g\in\mathcal{M}_\kappa^5,\ \ g\star g^\dag=g^\dag\star g=\bbone  \},\label{NCsym}
\end{equation}
which can be viewed as the noncommutative analog of the $U(1)$ group.\\

Now, recall that the twisted connection is defined \cite{MW20} as a map $\nabla_{X_\mu}:\mathbb{E}\to\mathbb{E}$, for any $X_\mu\in\mathfrak{D}_\gamma$, fulfilling the following properties:
\begin{eqnarray}
\nabla_{X_\mu+X^\prime_\mu}(m)&=&\nabla_{X_\mu}(m)+\nabla_{X^\prime_\mu}(m)\label{sigtaucon1},\\
\nabla_{z.X_\mu}(m)&=&\nabla_{X_\mu}(m)\star z\label{sigtaucon2},\\
\nabla_{X_{\mu}}(m\star f)&=&\nabla_{X_{\mu}}(m)\star(\mathcal{E}^{\gamma}\triangleright f)+(\mathcal{E}^{\gamma+1}\triangleright m)\star X_{\mu}(f),\label{sigtauconbis1}
\end{eqnarray}
for any $m\in\mathbb{E}\simeq\mathcal{M}_\kappa^5$, $X_\mu,X^\prime_\mu\in\mathfrak{D}_\gamma$, $z\in \mathcal{Z}(\mathcal{M }_\kappa^5)$ (the center of $\mathcal{M}_\kappa^5$), $f\in\mathcal{M}_\kappa^5$. In \eqref{sigtauconbis1} the factor $(\mathcal{E}^{\gamma+1}\triangleright m)$ in the 2nd term must be understood as a morphism $\tilde{\beta}:\mathbb{E}\to\mathbb{E}$ acting on the module as $\tilde{\beta}(m)=\mathcal{E}^{\gamma+1}\triangleright m$ for any $m$ in $\mathbb{E}\simeq\mathcal{M}_\kappa^5$. \\
Set 
\begin{equation}
A_\mu:=\nabla_{X_{\mu}}(\bbone), \ \ \nabla_\mu:=\nabla_{X_\mu}\label{cestamiou}.
\end{equation}
Then, observe that the noncommutative analog of the "gauge potential" $A_\mu$ defined above and verifying
\begin{equation}
\nabla_\mu(f)=A_\mu\star \mathcal{E}^\gamma\triangleright f+X_\mu(f)
\end{equation}
obtained by setting $m=\bbone$ in \eqref{sigtauconbis1}, does not generally satisfy the usual relation $A_\mu^\dag=A_\mu$. \\
This is a mere consequence of the fact that the $X_\mu$'s are twisted and are {\it{not}} real derivations. Indeed, one has
\begin{equation}
(X_\mu(f))^\dag=-\mathcal{E}^{-2\gamma-1}\triangleright (X_\mu(f^\dag))\ne X_\mu(f^\dag). \label{laformule}
\end{equation}
\\
In fact, one finds after standard algebraic calculations that $A_\mu$ satisfies
\begin{equation}
A_\mu=\mathcal{E}^{2\gamma+1}\triangleright A_\mu^\dag\label{twistedhermit},
\end{equation}
together with the following twisted Hermiticity condition for the connection
\begin{equation}
h(\mathcal{E}^{-2\gamma-1}\triangleright i\nabla_{X_{\mu}}(m_1),\mathcal{E}^{\gamma}\triangleright m_2)+h(\mathcal{E}^{-\gamma-1}\triangleright m_1,i\nabla_{X_{\mu}}(m_2))=iX_\mu h(m_1,m_2)\label{cledag}
\end{equation}
which holds true for any $X_\mu\in\mathfrak{D}_\gamma$, $m_1,m_2\in\mathcal{M}_\kappa^5$.\\
Note that somewhat similar deformed Hermiticity condition for noncommutative connections also appeared within the framework of $\varepsilon$-derivations giving rise to the notion of $\varepsilon$-connections \cite{epsilonconnect}.\\

It is known that Hermitian connections play a central role in the physics described by (commutative) Yang-Mills theories. The present situation obviously deals with a noncommutative analog of a $U(1)$ Yang-Mills theory. Accordingly, we assume from now on that the twisted Hermitian condition for the connection \eqref{twistedhermit} holds true. To simplify the analysis, we will further {\it{assume}} that
\begin{equation}
\gamma=0\label{plusgamma},
\end{equation}
which will not alter the conclusions of this paper.\\

It will be useful in the sequel to introduce the 1-form connection $A$ and its associated curvature 2-form $F$. The main properties of the corresponding noncommutative differential calculus and the related notations introduced in \cite{MW20} are collected for convenience in the appendix \ref{appendixA}. This provides a convenient formalism to deal with the BRST symmetry. \\
We will introduce below the material used in the ensuing analysis. The curvature 2-form $F\in\Omega^2(\mathfrak{D}_0)$  is easily found to be given by
\begin{equation}
F={\bf{d}}A+((\mathcal{E}\triangleright A)\times A)\label{form-curv}
\end{equation}
where $A\in\Omega^1(\mathfrak{D}_0)$, $\times$ denotes the associative product of forms and ${\bf{d}}$ is the twisted differential with 
\begin{equation}
{\bf{d}}^2=0
\end{equation}
satisfying the twisted Leibniz rule
\begin{equation}
{\bf{d}}(\omega\times\eta)={\bf{d}}\omega\times\eta+(-1)^{\delta(\omega)}(\mathcal{E}\triangleright\omega)\times{\bf{d}}\eta\label{leib-differ}
\end{equation}
for any $\omega,\ \eta\in\Omega^\bullet$ where $\delta(\omega)$ is the form degree of $\omega$. Recall that one has $A(X_\mu)=A_\mu$ and $F(X_\mu,X_\nu)=F_{\mu\nu}$. \\
In \eqref{form-curv}, \eqref{leib-differ} and in the sequel of the discussion, the action of $\mathcal{E}$ on the forms must be understood as $(\mathcal{E}\triangleright\omega)(X_1,...,X_p)=(\mathcal{E}\triangleright(\omega(X_1,...,X_p)))$ for any $\omega\in\Omega^\bullet$, i.e. $\mathcal{E}$ acts on the "components of the forms".\\ 

One can verify that the classical action is invariant under the gauge transformations
\begin{eqnarray}
A_\mu^g&=&(\mathcal{E}\triangleright g^\dag)\star A_\mu\star g+ (\mathcal{E}\triangleright g^\dag)\star X_\mu(g),\label{transamu}\\
F_{\mu\nu}^g&=&(\mathcal{E}^2\triangleright g^\dag)\star F_{\mu\nu}\star g \label{transfmunu},
\end{eqnarray}
or equivalently on the connection 1-form and curvature 2-form
\begin{eqnarray}
A^g&=&(\mathcal{E}\triangleright g^\dag)\times A\times g+(\mathcal{E}\triangleright g^\dag)\times {\bf{d}}g\label{transformA},\\
F^g&=&(\mathcal{E}^2\triangleright g^\dag)\times F\times g \label{transformF}
\end{eqnarray}
which hold true for any $g$ of the gauge group $\mathcal{U}$.\\

\subsection{BRST symmetry and the gauge-fixed action.}\label{section22}
The BRST symmetry associated with \eqref{transamu}-\eqref{transformF} is defined by the following structure equations \cite{Mathieu:2021mxl}
\begin{eqnarray}
s_0A&=&-{\bf{d}}C-(\mathcal{E}\triangleright C)\times A-A\times C\label{sA},\\
s_0C&=&-C\times C\label{sc},
\end{eqnarray}
so that the BRST transformation of the curvature 2-form is
\begin{equation}
s_0F=F\times C-(\mathcal{E}^2\triangleright C)\times F\label{sF},
\end{equation}
and one has $s_0^2=0$. Here, $C$ is the Fadeev-Popov ghost, a real-valued field with ghost number $+1$ while $s_0$ is the Slavnov operation associated with the gauge transformations \eqref{transamu}, \eqref{transfmunu}, whose action on any field increases its ghost number by $+1$. The relevant technical materials needed in the ensuing discussion are collected in the appendix \ref{appendixA}.\\

The transformations of the components are easily found to be given by
\begin{eqnarray}
s_0A_\mu&=&X_\mu(C)-(\mathcal{E}\triangleright C)\star A_\mu+A_\mu\star  C\label{samu},\\
s_0C&=&-C\star C\label{decadix},\\
s_0F_{\mu\nu}&=&F_{\mu\nu}\star C-(\mathcal{E}^2\triangleright C)\star F_{\mu\nu}\label{sFmunu}
\end{eqnarray}
upon using $(A\times C)(X_\mu)=-A_\mu\star C$, $(C\times A)(X_\mu)=C\star A_\mu$, $(dC)(X_\mu)=-X_\mu(C)$. Furthermore, one
can verify that
\begin{equation}
s_0S_{cl}=0\label{invarclass}.
\end{equation}
\\
Recall that a suitable framework encompassing the differential calculus and the BRST symmetry is obtained by introducing a bigraded differential calculus. For more mathematical details relevant to the present situation, see \cite{Mathieu:2021mxl}. \\
In particular, $s_0$ acts as an {\it{untwisted}} graded derivation. Namely, it satisfies the following Leibniz rule
\begin{equation}
s_0(\omega\times\eta)=s_0\omega\times\eta+(-1)^{\vert\omega\vert}\omega\times s_0\eta
\end{equation}
for any bigraded forms $\omega, \eta\in{\widehat{\Omega}}=\bigoplus_{p,q}\Omega^{p,q}(\mathfrak{D}_0)$, where $\vert\omega\vert$ denotes the total degree of $\omega$ defined as the sum of the form degree $\delta(\omega)$ and the ghost number of $\omega$. Accordingly, the Leibniz rule for ${\bf{d}}$ \eqref{leib-differ} still holds for any bigraded forms  $\omega, \eta\in{\widehat{\Omega}}$ with however $\delta(\omega)$ replaced by $\vert\omega\vert$. \\
Notice by the way that one should have $C=C^\dag$, stemming from the fact that $s_0$ can be viewed as a Grassmann version of the infinitesimal gauge transformations. This, combined with the Hermiticity relation \eqref{twistedhermit} together with \eqref{sA} yields $s_0(A^\dag)=(s_0(A))^\dag$.\\

The BRST operation $s_0$ generates the functional Slavnov identity which serves to control the UV behavior of the action $S_{cl}$ after its gauge-fixing. This latter is obtained by adding a BRST-exact term. A convenient gauge-fixing is given by 
\begin{equation}
S=S_{cl}+s_0\int d^5x\ ({\overline{C}}^\dag\star(\mathcal{E}^{-4}\triangleright X_\mu(A_\mu))\label{gaugefixing},
\end{equation}
supplementing the BRST structure equations \eqref{samu}-\eqref{sFmunu} by
\begin{eqnarray}
s_0\overline{C}^\dag&=&b^\dag,\label{scbar}\\
s_0b^\dag&=0&\label{sb}
\end {eqnarray}
where $\overline{C}$ (resp. $b$) is the antighost (resp. St\"uckelberg) real-valued field with ghost number $-1$ (resp. $0$). A simple calculation yields
\begin{equation}
S=S_{cl}+\int d^5x\ \big(b.X_\mu(A_\mu)+\overline{C}.X^2(C)-\overline{C}.X_\mu((\mathcal{E}\triangleright C)\star A_\mu-A_\mu\star  C)\big), \label{actionfinale}
\end{equation}
where we set $X^2=X^\mu X_\mu$ and we used the twisted trace property \eqref{twistrace} for $\int d^5x$ together with the useful identity
\begin{equation}
\int d^5x\ (f\star g^\dag)(x)=\int d^5x\ f(x).\bar{g}(x)\label{identity}.
\end{equation}
\\

The quadratic part of the action defines the kinetic terms for the gauge potential and the ghosts. It is given by
\begin{equation}
S_{kin}=\int d^5x\ \frac{1}{2}A_\mu(X^2\delta_{\mu\nu} -X_\mu X_\nu)A_\nu+\overline{C}.X^2(C)\label{action-kinetic},
\end{equation}
where the term involving $A_\mu$ is easily obtained upon using $\langle a,X_\mu(b )\rangle=\langle X_\mu(a),b \rangle$ for any $a,b\in\mathcal{M}_\kappa^5$ where 
\begin{equation}
\langle a, b\rangle:=\int d^5x\ a^\dag\star b\label{hermit-product} 
\end{equation}
is the hermitian product introduced in \cite{PW2018}. \\

Performing the functional integration over the $b$ field in the generating functional of the Green functions enforces the constraint $X_\mu(A_\mu)=0$ everywhere in the gauge-fixed action \eqref{actionfinale}. Accordingly, the kinetic terms for the gauge potential and the ghosts are respectively given by
\begin{eqnarray}
S_{kin}(A)&=&\frac{1}{2}\int d^5xd^5y\ A_\mu(x)K(x-y)A_\mu(y)\label{kinet-amiou},\\
S_{kin}(\overline{C},C)&=&\int d^5pd^5q\ \overline{C}(x)K(x-y)C(y)\label{kinet-ghost},
\end{eqnarray}
with
\begin{equation}
K(x-y)=\int d^5p\ e^{ip(x-y)}T(p)\label{KXY}
\end{equation}
in which
\begin{equation}
T(p)=(\vec{p}\ ^2+\kappa^2(1-e^{-{p_0}/{\kappa}})^2)\label{operat-kinet}.
\end{equation}
Notice that $K(x-y)\ne K(y-x)$ which simply comes from the exponential $e^{-{p_0}/{\kappa}}$ in the second term of \eqref{operat-kinet} so that $T(p)\ne T(-p)$. It is convenient to re-express \eqref{kinet-amiou} as
\begin{equation}
S_{kin}(A)=\frac{1}{2}\int d^5xd^5y\ A_\mu(x)S(x-y)A_\mu(y),\label{s1}
\end{equation}
with
\begin{equation}
S(x-y)=\frac{1}{2}\int d^5p\ e^{ip(x-y)}(T(p)+T(-p)).\label{s2}
\end{equation}
\\

The trilinear gauge potential interaction can be written as
\begin{equation}
S_{AAA}=\langle (X_\mu A_\nu-X_\nu A_\mu)^\dag,((\mathcal{E}\triangleright A_\mu)\star A_\nu-(\mathcal{E}\triangleright A_\nu)\star  A_\mu)^\dag\rangle + \textrm{c.c}\label{vertex-AAA}.
\end{equation}
The corresponding vertex function can be read off from (we use the obvious notation for the momenta: $p=(p_0,\vec{p})$)
\begin{equation}
S_{AAA}=\int d^5pd^5qd^5r\ A_\mu(p) A_\nu(q)A_\nu(r)V^A_\mu(p,q,r)\delta(p_0+q_0+r_0)\label{vert-A}
\end{equation}
with
\begin{equation}
V^A_\mu(p,q,r)=Q_\mu(q_0,\vec{q})(e^{-3p_0/\kappa}\delta(\vec{p}e^{-p_0/\kappa}+\vec{q}e^{-p_0/\kappa}+\vec{r})-e^{-3r_0/\kappa}\delta(\vec{p}+\vec{q}e^{-r_0/\kappa}+\vec{r}e^{-r_0/\kappa}))\label{vertexgauge1},
\end{equation}
where
\begin{equation}
Q_\mu(q_0,\vec{q})=(\kappa(1-e^{-q_0/\kappa}),q_i)\label{grandq}.
\end{equation}
\\

The trilinear gauge-ghost interaction can be cast into the form
\begin{equation}
S_{AC\overline{C}}=\int d^5x\ \overline{C}(x).\big((\mathcal{E}\triangleright A_\mu)\star X_\mu(C)-(\mathcal{E}\triangleright X_\mu(C))\star A_\mu \big)
\end{equation}
where use has been made of the gauge function $X_\mu A_\mu=0$. After some algebra, it can be put into the form (again we use the obvious notation for the momenta: $p=(p_0,\vec{p})$)
\begin{equation}
S_{AC\overline{C}}=\int d^5pd^5qd^5r\ \overline{C}(p)A_\mu(q)C(r)V^{\phi\pi}_\mu(p,q,r)\delta(p_0+q_0+r_0)\label{vertexghost1},
\end{equation}
with
\begin{equation}
V^{\phi\pi}_\mu(p,q,r)=Q_\mu(r_0,\vec{r})(e^{-q_0/\kappa}\delta(\vec{p}+\vec{q}+\vec{r}e^{-q_0/\kappa})-e^{-r_0/\kappa}\delta(\vec{p}+\vec{r}+\vec{q}e^{-r_0/\kappa}))\label{vertex-ghost-new}
\end{equation}
where $Q_\mu$ is still defined by \eqref{grandq}.\\

One easily verifies that both vertex functions vanish in the commutative limit, which is consistent with the usual electrodynamics for which ghosts decouple.\\

\section{Computation of the 1-point function.}\label{section3}
\subsection{Perturbative set-up.}\label{section31}
Introduce the generating functional of the connected Green functions $W(J,\overline{\eta},\eta)$ defined by
\begin{equation}
e^{W(J,\overline{\eta},\eta)}=\int dAd\overline{C}dCe^{-(S+S_{sources})}\label{W}
\end{equation}
where $S$ is given by \eqref{actionfinale} and $J,\overline{\eta},\eta$ are as usual sources to be defined just below. The part of the action relevant for the ensuing calculation involves the kinetic part together with the cubic vertices. We write the source term as
\begin{equation}
S_{sources}=\int d^5x\ A_\mu(x)J_\mu(x)+\overline{\eta}(x)C(x)+\overline{C}(x)\eta(x)\label{source}
\end{equation}
where $J$, $\overline{\eta}$, $\eta$ are sources associated respectively to $A_\mu$, $C$, $\overline{C}$ with respective ghost numbers $0,-1,1$. \\

Recall that functional derivatives obey obvious graded Leibniz rules with grading defined by the ghost number. Besides, Green functions involving ghost fields are generated by the action on \eqref{W} of functional derivatives $\frac{\delta}{\delta{\overline{\eta}}}$ and $\frac{\delta}{\delta \eta}$, acting respectively from left and right. Namely, one has
\begin{eqnarray}
\frac{\delta}{\delta{\overline{\eta}}(x)}e^{\int d^5y\ \overline{\eta}(y)C(y)+\overline{C}(y)\eta(y)}&=&C(x)e^{\int d^5y\ \overline{\eta}(y)C(y)+\overline{C}(y)\eta(y)}\label{delta1},\\
\frac{\delta}{\delta \eta(x)}e^{\int d^5y\ \overline{\eta}(y)C(y)+\overline{C}(y)\eta(y)}&=&e^{\int d^5y\ \overline{\eta}(y)C(y)+\overline{C}(y)\eta(y)}\overline{C}(x)\label{delta2}.
\end{eqnarray}
\\
The perturbative expansion is generated by the following functional relation
\begin{equation}
W(J,\overline{\eta},\eta)=W_0(J,\overline{\eta},\eta)+\ln\big(1+e^{-W_0(J,\overline{\eta},\eta)}[e^{-S_{int}}-1]e^{W_0(J,\overline{\eta},\eta)}\big)\label{generateur}
\end{equation}
up to an unessential additive constant, where $W_0(J,\overline{\eta},\eta)$ denotes the free generating functional of the connected Green functions. It is given by
\begin{equation}
W_0(J,\overline{\eta},\eta)=\int d^5xd^5y\ (\frac{1}{2}J_\mu(x)S^{-1}(x-y)J_\mu(y)+\overline{\eta}(x)K^{-1}(x-y)\eta(y))\label{Wzero},
\end{equation}
with
\begin{equation}
S^{-1}(x-y)={2}\int d^5p\ e^{ip(x-y)}(T(p)+T(-p))^{-1}\label{Ksymet},
\end{equation}
\begin{equation}
K^{-1}(x-y)=\int d^5p\ e^{ip(x-y)}T^{-1}(p)
\end{equation}
where $T(p)$ still given by \eqref{operat-kinet}.\\
In \eqref{generateur}, one has
\begin{equation}
S_{int}=S_{int}(\frac{\delta}{\delta J},\frac{\delta}{\delta{\overline{\eta}}},\frac{\delta}{\delta\eta})=S_{AAA}(\frac{\delta}{\delta J})+S_{AC\overline{C}}(\frac{\delta}{\delta J},\frac{\delta}{\delta{\overline{\eta}}},\frac{\delta}{\delta\eta}),\label{esint}
\end{equation}
which is obtained by replacing each field in $S_{AAA}$ and $S_{AC\overline{C}}$, respectively \eqref{vertexgauge1}, \eqref{vertexghost1}, by its associated functional derivative. We did not explicitly write the quartic interaction since it will not generate contributions ot the tadpole 1-point function. \\

\subsection{The tadpole at one-loop.}\label{section32}

The one-loop contribution to the 1-point tadpole function can be extracted from
\begin{equation}
W(J,\overline{\eta},\eta)=W_0(J,\overline{\eta},\eta)-e^{-W_0(J,\overline{\eta},\eta)}S_{int}(\frac{\delta}{\delta J}\frac{\delta}{\delta{\overline{\eta}}},\frac{\delta}{\delta\eta})e^{W_0(J,\overline{\eta},\eta)}\label{utile}.
\end{equation}
\\

The relevant part of $W(J,\overline{\eta},\eta)$ corresponding to the one-loop ghost contribution to the 1-point Green function for $A_\mu$, denoted hereafter by $W_1^{\phi\pi}(J)$, is obtained by simply combining \eqref{vertexghost1}, \eqref{vertex-ghost-new}, \eqref{Wzero}, \eqref{esint} with \eqref{utile}. A standard calculation yields
\begin{eqnarray}
W_1^{\phi\pi}(J)&=&\int d^5pd^5qd^5rd^5xd^5yd^5zd^5w\ e^{-i(px+qy+rz)}K^{-1}(z-x)\\ \nonumber
&\times&S^{-1}(y-w)J_\mu(w)\delta(p_0+q_0+r_0)V^{\phi\pi}_\mu(p,q,r)\label{W1G}
\end{eqnarray}
where $V^{\phi\pi}_\mu(p,q,r)$ is still given by \eqref{vertex-ghost-new}. By further making use of the Legendre transform 
$\frac{\delta W}{\delta J_\mu(x)}=A_\mu(x)$ which at the 1st order, relevant here, reduces to
\begin{equation}
A_\mu(x)=\int d^5y\ S^{-1}(x-y)J_\mu(y),\label{legendre}
\end{equation}
one readily derives from \eqref{W1G}, combined with $W_1^{\phi\pi}(J)$, the corresponding ghost contribution to the one-loop effective action, denoted by $\Gamma_1^{\phi\pi}(A)$. By taking into account the various delta functions occurring in $W_1^{\phi\pi}(J)$, we find that the expression reduces to
\begin{equation}
\Gamma_1^{\phi\pi}(A)=\int d^5q\ \delta(q)A_\mu(q)\mathcal{I}_\mu\label{tadpole-ghost},
\end{equation}
in which
\begin{equation}
\mathcal{I}_\mu=\int d^5s\ \frac{1-e^{3s_0/\kappa}}{\vec{s}^{\ 2}+\kappa^2(1-e^{-s_0/\kappa})^2}\times Q_\mu(s)\label{integrale-ghost}.
\end{equation}
From \eqref{integrale-ghost}, one immediately infers that the spatial components of $\mathcal{I}_\mu$ vanish since the following relation $\int d\vec{s}\ \frac{s_i}{\vec{s}^{\ 2}+M^2}=0$, $i=1,...,4$ ($M^2$ is some positive quantity) holds true. Hence, one concludes that the ghosts only contribute to the 1-point function for $A_0$. One obtains finally
\begin{equation}
\Gamma_1^{\phi\pi}(A)=\int d^5x\ A_0(x)\mathcal{I}_0\label{tadpole-ghost-final}.
\end{equation}
Notice that one can easily verify that $\lim_{\kappa\to\infty}\mathcal{I}_0=0$ hence signaling the vanishing of $\Gamma_1^{\phi\pi}(A)$ at the commutative limit as expected.\\

The relevant part of $W(J,\overline{\eta},\eta)$ corresponding to the one-loop $A_\mu$ contribution to the 1-point Green function for $A_\mu$, denoted hereafter by $W_1^{A}(J)$ can be computed in a similar way. We find
\begin{eqnarray}
W_1^{A}(J)&=&\int d^5pd^5qd^5rd^5xd^5yd^5zd^5w\ e^{-i(px+qy+rz)}\delta(p_0+q_0+r_0)\\ \nonumber
&\times&V^A_\mu(p,q,r)J_\mu(w)\big(5S^{-1}(x-w)S^{-1}(y-z) \\\nonumber
&+& S^{-1}(y-w)S^{-1}(z-x)+S^{-1}(z-w)S^{-1}(x-y)   \big)\label{W1A}.
\end{eqnarray}
Upon using \eqref{legendre},  the corresponding contribution of the gauge potential to the one-loop effective action, denoted by $\Gamma_1^{\phi\pi}(A)$ can be written as a sum of 3 terms
\begin{equation}
\Gamma_1^A(A):=\Gamma_{11}(A)+\Gamma_{12}(A)+\Gamma_{13}(A)\label{gammaAAA}
\end{equation}
with
\begin{eqnarray}
\Gamma_{11}^A(A)&=&\int d\theta_\mu\ S^{-1}(z-x)A_\mu(y)\label{gamma11}\\
\Gamma_{12}^A(A)&=&\int d\theta_\mu\ S^{-1}(x-y)A_\mu(z)\label{gamma12}\\
\Gamma_{13}^A(A)&=&5\int d\theta_\mu\ S^{-1}(y-z)A_\mu(x)\label{gamma13}
\end{eqnarray}
where 
\begin{equation}
d\theta_\mu:=d^5pd^5qd^5rd^5xd^5yd^5z\ \ \ e^{-i(px+qy+rz)}\delta(p_0+q_0+r_0)V^A_\mu(p,q,r)\label{dtheta}.
\end{equation}
The first contribution $\Gamma_{11}^A(A)$ vanishes. Indeed, from simple manipulations of the various delta functions occurring in $\Gamma_{11}^A(A)$, one easily realizes that the contribution to $A_0$ vanishes, due to the fact basically that one factor $\sim\delta(q_0)Q_0(q)$ appears in the expression which upon integrating over $q_0$ gives 0. For the spatial contributions, one arrives at
\begin{equation}
\Gamma_{11}^A(A)\sim\int d^5qds_0d^5yA_i(y)q_ie^{-iqy}\delta(q_0)\big( S^{-1}(-\vec{q}\Sigma(s_0),q_0)-S^{-1}(\vec{q}\Sigma(s_0),q_0)  \big)=0
\end{equation}
where the last equality comes from the fact that $S^{-1}(p)$ is an even function of $\vec{p}$. Hence
\begin{equation}
\Gamma_{11}^A(A)=0.\label{tadzero}
\end{equation}
The second contribution $\Gamma_{12}^A(A)$ can be cast into the form
\begin{equation}
\Gamma_{12}^A(A)=\int d^5x A_\mu(x)\mathcal{J}_\mu\label{gamma-pasnul}
\end{equation}
with
\begin{equation}
\mathcal{J}_\mu=\int d^5s\ 2Q_\mu(s)\frac{e^{3s_0/\kappa}-1}{T(s)+T(-s)}.\label{cjmu}
\end{equation}
By observing that $\mathcal{J}_i=0$, $i=1,...,4$, one concludes that 
\begin{equation}
\Gamma_{12}^A(A)=\int d^5x\ A_0\mathcal{J}_0\label{gamma-pasnul0}.
\end{equation}
Finally, the last contribution $\Gamma_{13}^A(A)$ yields
\begin{equation}
\Gamma_{13}^A(A)=-5\int d^5x A_\mu(x)\mathcal{J}_\mu=-5\int d^5x\ A_0\mathcal{J}_0.\label{ouf}
\end{equation}

From \eqref{tadzero}, \eqref{gamma-pasnul0}, \eqref{ouf} and \eqref{tadpole-ghost-final}, one concludes that the contribution of the 1-point function for the gauge potential is given by
\begin{equation}
\Gamma_1(A)=\int d^5x\ A_0(x)(\mathcal{I}_0-4\mathcal{J}_0)\label{Zetadpole},
\end{equation}
which, in view of \eqref{integrale-ghost} and \eqref{cjmu} is non zero.\\

Hence, only the time component $A_0$ of the gauge potential has a non-zero 1-point function at the one-loop order.

\section{Discussion.}\label{section4}

We have considered a gauge theory on the (necessarily 5-d) $\kappa$-Minkowski space which can be viewed as the noncommutative analog of a $U(1)$ gauge theory. First, we have shown in Subsection \ref{section21} that the Hermiticity condition obeyed by the gauge potential $A_\mu$ is twisted. Then, we have carried out a first exploration of the quantum properties of this gauge theory suitably gauge-fixed using the twisted BRST symmetry framework elaborated in \cite{Mathieu:2021mxl}. Assuming that $A_\mu$ is real-valued and working at the one-loop order, we have find that the gauge-fixed theory gives rise to a non-vanishing tadpole for the $A_\mu$ corresponding to the occurrence in the 1-loop effective action of a term linear in $A_\mu$ of the form
\begin{equation}
\Gamma_1(A)\sim\int d^5x\ \sigma_\mu A_\mu(x)=\int d^5x\ K(\kappa)A_0(x)\label{resulfinal},
\end{equation}
where the diverging quantity $K(\kappa)$ can be read of from \eqref{Zetadpole} and must then be suitably regularized.\\

Had we have relaxed the assumption on $A_\mu$ to be real-valued{\footnote{The commutative limit would however involve two gauge potentials.}}, thus starting from a complex-valued $A_\mu$, then we would have find again a non-vanishing tadpole. This can be verified from a computation similar to the one presented in Section \ref{section3}. Indeed, to compute the 1-point contribution for $A_\mu$, simply replace everywhere $S(x-y)$ defined in \eqref{s1}, \eqref{s2} by $K(x-y)$ \eqref{KXY}, \eqref{operat-kinet}, with the relevant trilinear vertex now given by
\begin{eqnarray}
 S_{AAA}&=&\int d^5pd^5qd^5r\ \overline{A}_\mu(p) A_\nu(q)\overline{A}_\nu(r)Q_\mu(q)V^A_\mu(p,q,r)\delta(p_0-q_0+r_0)\nonumber\\
&\times&\big(e^{3p_0/\kappa} \delta(\vec{p}e^{p_0/\kappa}-\vec{q}e^{p_0/\kappa}+\vec{r})- e^{3r_0/\kappa}\delta(\vec{p}-\vec{q}e^{r_0/\kappa}+\vec{r}e^{r_0/\kappa})\big),
\end{eqnarray}
and the gauge-ghost vertex unchanged, leading to $\Gamma_1(A,\overline{A})\sim \int d^5x\ A_0(x)\mathcal{I}_0$. One proceeds similarly for the $\overline{A}_\mu$ contribution. \\
Notice that the inclusion of fermions{\footnote{Recall that the corresponding commutative limit would lead to a nonrenormalisable model. }}, obtained by supplementing the action with the following gauge invariant coupling 
\begin{equation}
S_F=\int d^5x\ (\psi^\dag\gamma^0\star\mathcal{E}^{-1}\gamma^\mu\nabla_\mu\psi)(x),
\end{equation}
with $\nabla_\mu\psi=A_\mu\star \psi+X_\mu(\psi)$, does not change the conclusion. This simply stems from the fact that the corresponding fermionic contribution to the 1-point function identically vanishes as in the commutative case, being proportional to the trace of a single gamma matrice, which can be easily verified by using the fermion propagator whose expression is given by $K^{-1}_F(x-y)=\int d^5x\ e^{ip(x-y)}\gamma^\mu Q_\mu(p)T^{-1}(p)$.\\

The appearance of a non-zero tadpole at the 1-loop order has already been evidenced in various classes of gauge theories on quantum spaces. For instance, this shows up in the massless gauge theory on $\mathbb{R}^3_\lambda$, a deformation of the 3-d space \cite{geviwa}. \\
In the same way, a tadpole appears in a familly of gauge matrix models on the Moyal plane $\mathbb{R}^2_\theta$ \cite{vign-sym}. Recall that in these gauge models, the relevant field variable is a tensor form, sometimes called the covariant coordinate, which is the difference of two form-connections, one of them having a distinguished status{\footnote{This latter is the canonical gauge invariant connection rigidely linked with the coordinates of the Moyal space. For more details, see e.g. \cite{cawa}, \cite{wal-moyal1}.}}. The corresponding (classical) action functional is quadratic and quartic in the tensor form but must be expanded around a symmetric vacuum \cite{wal-moyal2} in order to obtain a dynamically non-trivial model, which generates an additional cubic vertex upon expansion. This results in a non-zero tadpole so that a term linear in $A_\mu$, albeit absent at the classical order, is induced by quantum fluctuations in the 1-loop effective action. \\
Note that in both cases, the relevant BRST symmetry is untwisted. Note also that the analysis in \cite{geviwa} is based on a temporal gauge, giving rise to a non vanishing tadpole. We will discuss more closely the consequences implied by the use of such a temporal gauge at the end of this section.\\

The complicated vacuum structure \cite{wal-moyal2} of the gauge matrix models on 4-d Moyal spaces $\mathbb{R}^4_\theta$ \cite{grosse11}, \cite{wal-moyal1} forbids, so far, their complete exploration at the quantum level. Note however that a class of intensively studied gauge theories on $\mathbb{R}^4_\theta$ \cite{gauge-moy1}-\cite{gauge-moy11}, bearing a formal similarity with the (commutative) Yang-Mills theories, do not produce 1-loop tadpoles. This stems from the mere algebraic structure of the trilinear gauge potential vertices which produces automatically vanishing tadpole contributions. Unfortunately, these noncommutative gauge theories suffer from UV/IR mixing \cite{uvir1}, which thus likely precludes the achievement of their perturbative renomalisability{\footnote{Note that an interesting interpretation of the UV/IR mixing in term of an induced gravity action has been presented in e.g. \cite{harold1}, \cite{harold2}. It takes place in a matrix formulation of the gauge theories on Moyal space.}} . \\

At this stage, some comments are in order. \\

A non-vanishing tadpole is linked with a non-vanishing 1-point function, i.e. vacuum expectation value (vev), for the gauge potential, say $\langle A_\mu \rangle\ne 0$, which may have some noteworthy consequences as gauge symmetry breaking as well as Lorentz 
symmetry breaking may occur. In the present situation, the occurrence of the extra term $\Gamma_1(A)$ in the 1-loop effective action $\Gamma_{\mathrm{eff}}$ signals that the classical vacuum of the theory is not stable against quantum fluctuations. This can be traced back to the fact that the term linear in $A_\mu$ induced by "radiative corrections" obviously prevents the classical vacuum configuration characterized by $A_\mu=0$ to be an extremal point of $\Gamma_{\mathrm{eff}}$ (i.e. it does no longer verify the equations of motion related to $\Gamma_{\mathrm{eff}}$). Note that $\Gamma_1(A)$ is not gauge (BRST) invariant and that it seems unlikely possible to balance its gauge variation by another variation of some other higher order terms involved in $\Gamma_{\mathrm{eff}}$ suggesting that the classical symmetry is broken. \\
Getting rid of the linear term in $A_\mu$ is achieved as usual by expanding $\Gamma_{\mathrm{eff}}$ around the new vacuum $\tilde{A}_\mu${\footnote{solving the equations of motion related to $\Gamma_{\mathrm{eff }}$.}}, i.e. setting $A_\mu=\tilde{A}_\mu+\alpha_\mu$ in $\Gamma_{\mathrm{eff}}$ where the new field variable is $\alpha_\mu$, while the resulting (background-)symmetry of the expanded action should be presumably obtained from the BRST operation \eqref{samu}, \eqref{decadix} combined with the field expansion. Its full characterization would require to carry out the complete 1-loop renormalization which is beyond the scope of this paper.\\

 Besides, we note that having $\langle A_\mu \rangle\ne 0$ from \eqref{Zetadpole} is reminiscent of some instance in which a Lorentz symmetry breaking does occur. Among the numerous models describing possible Lorentz violations (for a review, see for instance \cite{cenous1} and references therein), some effective models leading to a spontaneous Lorentz symmetry breaking \cite{Kost} have been considered for some time, see e.g. \cite{Kost1}-\cite{chk11}. In these models, defined either in flat space or coupled to a gravitational field, a non-zero vev of a vector field, mostly obtained through the introduction of a suitable potential at the classical level, triggers the Lorentz symmetry breaking which may possibly exhibit some interplay with a kind of generation of a gauge symmetry \cite{hbnielsen}. \\
 The gauge theory on $\kappa$-Minkowski space considered in this paper bears at a first sight somewhat similar gross features with the models mentioned above. However, it differs in some respects. Indeed, one starts with a non-zero vev of a gauge potential, instead of a simple vector field. This non-zero value is induced by radiative corrections and is not present at the classical level. Besides, the gauge theory under consideration is non local while the models mentioned above are essentially local. Nevertheless, note that the relevant field variable $\alpha_\mu$ obtained after achieving the expansion of the effective action around the new vacuum $\tilde{A}_\mu$ is a vector field as being equal to the difference of two connections.\\
\vfill\eject
It is instructive to notice that the expression \eqref{resulfinal} actually depends on the gauge choice. This gauge dependence can be made more apparent by choosing a one-parameter family of "non-covariant" gauges involving the temporal/Weyl gauge $A_0=0$ for a special value of the parameter. It is worth recalling that this latter gauge choice generates some difficulties when performed within commutative gauge theories, stemming, for instance in QED, from the peculiar momentum dependence of the gauge-fixed photon propagator. Other non-covariant gauges are as well not free from difficulties. However, some of these difficulties can be (partly) overcome or circumvented. For technical details on non-covariant gauges, see e.g. \cite{leib}, \cite{basset}.\\

For our present purpose, it is convenient to start, instead of \eqref{gaugefixing}, from the following gauge-fixing action 
\begin{equation}
S_{GF}=s_0\int d^5x\ \left( {\overline{C}}^\dag\star\mathcal{E}^{-4}\triangleright  \left(\frac{\lambda}{4}b-A_0 \right) \right)
=\int d^5x\ {\overline{C}} \left( s_0A_0 + b \left(\frac{\lambda}{4}b-A_0 \right) \right)
\end{equation}
where $\lambda$ is a real parameter. The functional integration over the St\"uckelberg field $b$ yields 
\begin{equation}
S_{GF}=\int d^5x\ -\frac{1}{\lambda}A_0^2+{\overline{C}}X_0 C+{\overline{C}}(A_0\star C-\mathcal{E}\triangleright C \star A_0)\label{gf2}.
\end{equation}
By carrying out a computation similar to the one presented in Section \ref{section32}, one easily realizes that \eqref{gf2} gives rise to a non-zero contribution to a tadpole for $A_0$  of the form (to be suitably regularized) $\Gamma_1^{\phi\pi}(A_0)\sim\int d^5x A_0(x)\times(\int d^5k\frac{F(k_0)}{(1-e^{-k_0/\kappa})})$ where $F(k_0)$ is a function whose exact expression is not needed here. But this latter contribution cannot be balanced by the corresponding contribution
from the cubic self-interaction for the $A_\mu$ which is easily found to be of the form $\Gamma_1^A(A_0)\sim\int d^5xA_0(x)\times(\int d^5k\ (Q_0(k)K_{00}(k)+Q_i(k)K_{0i}(k)))$ where $K_{00}(k)$ and $K_{0i}(k)$ are the components of the gauge-fixed propagator for $A_\mu${\footnote{$K_{0i}(p)=\frac{\lambda}{Q_0(p_0)}p_i,\ \ K_{00}(p)=\lambda,\ \ K_{ij}(p)=\frac{1}{\hat{p}^2}(\delta_{ij}+(\lambda\hat{p}^2+1)\frac{p_ip_j}{Q_0(p_0)^2})$ where $\hat{p}^2=Q_0(p_0)^2+\vec{p}^2.$}}. Again, a suitable regularization is understood. In view of the expressions for $K_{00}(k)$ and $K_{0i}(k)$, one obtains a contribution proportional to the gauge parameter $\lambda$, namely $\Gamma_1^A(A_0)\sim\lambda\int d^5x A_0(x)J$, where $J$ is some constant. Hence, there is a non-vanishing tadpole for $A_0$ in this gauge,
\begin{equation}
\Gamma_1(A_0)\sim \Gamma_1^{\phi\pi}(A_0)+\lambda\int d^5x A_0(x)J.\label{ccp}
\end{equation}
Besides, there is no tadpole for the spatial components of the gauge potential. To see that, one computes the corresponding 1-point function $\Gamma_1(A_I)$ for a given component $I$. Simple algebraic manipulations yield $\Gamma_1(A_I)\sim\int d^5x\ A_I(x)\times(\int d^5k\ K_{I0}(k)Q_0(k)+ K_{Ij}(k)Q_j(k))$ in obvious notations. But $K_{I0}(k)Q_0(k)$ and $K_{Ij}(k)Q_j(k)$ are linear in the internal spatial momentum $k_i$ so that the corresponding integrals in $\Gamma^1(A_I)$ vanish. Hence
\begin{equation}
\Gamma_1(A_I)=0\label{decadix22}.
\end{equation}
The usual temporal gauge $A_0=0$ is obtained by taking the limit $\lambda\to 0$ at which the ghosts decouple so that $\Gamma_1^{\phi\pi}(A_0)=0$. One can verify that \eqref{decadix22} still holds true while the remaining contribution from the cubic gauge interaction also vanishes, simply because it is proportional to the gauge parameter $\lambda$ as it is apparent in \eqref{ccp}. \\

Summarising this last observation, no tadpole does occur when the temporal gauge is used for the gauge-fixing. Note however that one can check by inspection that the temporal gauge does not fix the $x_0$-dependant $\mathcal{U}$ gauge transformations \eqref{NCsym}, as it is the case for the commutative case, so that the corresponding gauge-fixed theory should support a residual gauge symmetry.\\
In view of the above discussion dealing with different natural choices for gauge conditions, it would be worth analysing the consequences of the (possibly) non-vanishing 1-point function for $A_\mu$ from the viewpoint of Lorentz and noncommutative $\mathcal{U}$ gauge symmetries \eqref{NCsym}. We will come back to these aspects in a forthcoming work.\\
\vskip 1 true cm

{\bf{Acknowledgements}}:  We thank the Action CA18108 QG-MM, "Quantum Gravity Phenomenology in the multi-messengers approach", from the European Cooperation in Science and Technology (COST). Ph. M. is supported by the NSF grant 1947155 and the JTF grant 61521. J.-C. W thanks P. Martinetti for various discussions on twisted structures in noncommutative geometry.

\appendix

\section{Twisted differential calculus.}\label{appendixA}

Let $\mathfrak{D}_\gamma$ denotes the set of twisted derivation defined in eqn. \eqref{xmiou} and satisfying the Leibniz rule \eqref{leibnitz}. One can verify that $[X_\mu,X_\nu]=X_\mu X_\nu-X_\nu X_\mu=0$ so that $\mathfrak{D}_\gamma$ is an abelian Lie algebra. Let $\mathcal{Z}(\mathcal{M }_\kappa^d)$ be the center of $\mathcal{M }_\kappa^d$. We denote by $\Omega^n(\mathfrak{D}_\gamma)$ the linear space of $n$-linear antisymmetric forms. Note that linearity of forms holds w.r.t. $\mathcal{Z}(\mathcal{M}_\kappa^d)$. \\
The twisted differential calculus based on $\mathfrak{D}_\gamma$ is an extension of the derivation-based differential calculus introduced a long time ago. See e.g. \cite{mdv} and references therein.\\

 In the present situation, $n$-forms are defined from $\Omega^n(\mathfrak{D}_\gamma)$. For any $\alpha\in\Omega^n(\mathfrak{D}_\gamma)$, one has $\alpha:\mathfrak{D}_\gamma\to\mathcal{M}_\kappa^d$ together with
\begin{equation}
\alpha(X_1,X_2,...,X_n)\in\mathcal{M}_\kappa^d,\ \ \alpha(X_1,X_2,...,X_n.z)=\alpha(X_1,X_2,...,X_n)\star z\label{formule2},
\end{equation}
for any $z$ in $\mathcal{Z}(\mathcal{M }_\kappa^d)$ and any $X_1,...,X_n\in\mathfrak{D}_\gamma$. \\
Define now the linear space $\Omega^\bullet:=\bigoplus_{n=0}^{d}\Omega^n(\mathfrak{D}_\gamma)$, with $\Omega^0(\mathfrak{D}_\gamma)=\mathcal{M}_\kappa^d$. Then $(\Omega^\bullet,\times)$ is an associative algebra where the product of forms is defined for any $\alpha\in\Omega^p(\mathfrak{D}_\gamma)$, $\beta\in\Omega^q(\mathfrak{D}_\gamma)$ by $\alpha\times\beta\in\Omega^{p+q}(\mathfrak{D}_\gamma)$ with
\begin{align}
\nonumber
&(\alpha\times\beta)(X_1,...,X_{p+q})\\
&\qquad\qquad=\frac{1}{p!q!}\sum_{s\in\mathfrak{S}(p+q)}(-1)^{\text{sign}(s)}\alpha(X_{s(1)},...,X_{s(p)})\star \beta(X_{s(p+1)},...,X_{s(q)})\label{ncwedge1}.
\end{align}
In \eqref{ncwedge1}, $\mathfrak{S}(p+q)$ is the symmetric group of a set of $p+q$ elements, $\text{sign}(s)$ is the signature of the permutation $s$. Notice that $\alpha\times\beta\ne(-1)^{\delta(\alpha)\delta(\beta)}\beta\times\alpha$ where $\delta({\alpha})$ is the degree of $\alpha$ .\\
Now the triple $(\Omega^\bullet,\times,{\bf{d}})$ is a graded differential algebra where the differential ${\bf{d}}$ satisfies ${\bf{d}}:\Omega^p(\mathfrak{D}_\gamma)\to\Omega^{p+1}(\mathfrak{D}_\gamma)$, $p=0,...,(d-1)$ and
\begin{equation}
\left({\bf{d}}\alpha\right)\left(X_1,X_2,...,X_{p+1}\right)
=\sum_{i=1}^{p+1}(-1)^{i+1}X_i\left(\alpha(X_1,...,\vee_i,...,X_{p+1})\right), \label{ncwedge2}
\end{equation}
where the symbol $\vee_i$ indicates the omission of $X_i$. The differential satisfies ${\bf{d}}^2=0$ and the following twisted Leibniz rule
\begin{equation}
{\bf{d}}(\alpha\times\beta)={\bf{d}}\alpha\times\mathcal{E}^\gamma(\beta)+(-1)^{\delta(\alpha)}\mathcal{E}^{1+\gamma}(\alpha)\times{\bf{d}}\beta\label{leibniz-form},
\end{equation}
where $\mathcal{E}^x(\alpha)$ is defined for any $x\in\mathbb{R}$ and any $\alpha\in\Omega^n(\mathfrak{D}_\gamma)$ by $\mathcal{E}^x(\alpha)\in\Omega^n(\mathfrak{D}_\gamma)$ with $\mathcal{E}^x(\alpha)(X_1,...,X_n)=\mathcal{E}^x\triangleright(\alpha(X_1,...,X_n))$.\\
The inclusion of the BRST operation $s_0$, i.e. the Slavnov operation, in the above framework amounts to introduce bigraded forms which carry a ghost number in addition of the form degree. One has now ${\bf{d}}:\Omega^{p,g}(\mathfrak{D}_\gamma)\to\Omega^{p+1,g}(\mathfrak{D}_\gamma)$ and 
$s_0:\Omega^{p,g}(\mathfrak{D}_\gamma)\to\Omega^{p,g+1}(\mathfrak{D}_\gamma)$ where $\Omega^{p,g}(\mathfrak{D}_\gamma)$ is the space of $p$ forms with ghost number $g$. The Slavnov operation $s_0$ acts as a graded but {\it{untwisted}} derivation with the Leibniz rule $s_0(\rho\times\eta)=s_0(\rho)\times \eta+(-1)^{\vert\rho\vert}\rho\times s_0(\eta)$ for any $\rho, \eta\in\widehat{\Omega}(\mathfrak{D}_\gamma)$ where $\vert\rho\vert=\delta(\rho)+g$ with $\widehat{\Omega}(\mathfrak{D}_\gamma)=\bigoplus_{p,g}\Omega^{p,g}(\mathfrak{D}_\gamma)$ while \eqref{leibniz-form} still holds with however $\delta(\alpha)$ replaced by $\vert\alpha\vert$. For more mathematical details, see \cite{Mathieu:2021mxl}.

\vskip 1 true cm

\end{document}